\documentclass[aps,pra,reprint,superscriptaddress,showpacs]{revtex4-1}
\usepackage{graphicx}
\begin{document}

\title{Zero field line in the magnetic spectra of negatively charged nitrogen-vacancy centers in diamond}

\author{S.V.~Anishchik}
\email[]{svan@kinetics.nsc.ru}
\affiliation{Voevodsky Institute of Chemical Kinetics and Combustion SB RAS, 630090, Novosibirsk, Russia}

\author{V.G.~Vins}
\affiliation{VinsDiam Ltd., Russkaya str., 43, 630058, Novosibirsk, Russia}

\author{A.P.~Yelisseyev}
\affiliation{V.S. Sobolev Institute of Geology and Mineralogy SB RAS, 630090, Novosibirsk, Russia}

\author{N.N.~Lukzen}
\affiliation{International Tomography Center SB RAS, 630090, Novosibirsk, Russia}
\affiliation{Novosibirsk State University, 630090, Novosibirsk, Russia}

\author{N.L.~Lavrik}
\affiliation{Voevodsky Institute of Chemical Kinetics and Combustion SB RAS, 630090, Novosibirsk, Russia}

\author{V.A.~Bagryansky}
\affiliation{Voevodsky Institute of Chemical Kinetics and Combustion SB RAS, 630090, Novosibirsk, Russia}


\begin{abstract}
The dependence of the luminescence of diamonds with negatively charged nitrogen-vacancy centers (NV$^-$) vs. applied magnetic field (magnetic spectrum) was studied. A narrow line in zero magnetic field was discovered. The properties of this line are considerably different from those of other narrow magnetic spectrum lines. Its magnitude is weakly dependent of the orientation of the single-crystal sample to the external magnetic field. This line is also observed in a powdered sample. The shape of the line changes greatly when excitation light polarization is varied. The magnitude of the line has a non-linear relation to excitation light intensity. For low intensities this dependence is close to a square law. To explain the mechanism giving rise to this line in the magnetic spectrum, we suggest a model based on the dipole-dipole interaction between different NV$^-$ centers.
\end{abstract}

\pacs{61.72.jn, 75.30.Hx, 78.55.-m, 81.05.ug}

\maketitle

\section{INTRODUCTION}

Negatively charged nitrogen-vacancy center (NV$^-$ center) is of great interest to researchers due to their unique properties \cite{Doherty2013}. It is a promising system for numerous applications, especially in quantum information processing \cite{Gruber1997,Wrachtrup2001,Jelezko2004,Childress2006,Wrachtrup2006,Hanson2006b,Gaebel2006,Santori2006o,
Wrachtrup2006,Waldermann2007,Maurer2012,VanderSar2012,Neumann2013,Dolde2013,Dolde2014,Pfaff2014}.

The ground state of the NV$^-$ centers is triplet and split depending on the spin projection on the symmetry axis. Ground energy term is the term with zero spin projection. The splitting between this term and the ones with projections +1 and -1 in zero magnetic field is $\sim$2.88~GHz.

A spin projection-selective intersystem crossing from the excited triplet state to the excited singlet one as well as from the ground singlet state to the ground triplet state leads to, firstly, a much higher quantum yield of luminescence when excited from the zero spin projection state rather than from the states with projections +1 or -1. Secondly, it leads to a non-equilibrium state of the NV$^-$ center after multiple absorption and emission of light, with a population of the zero spin projection state much higher than those of the states with spin projections +1 or -1. The latter effect is usually referred to as optically-induced spin polarization \cite{Loubser1977,Manson2006,Delaney2010} despite the fact that in this state the spin projection on any direction is zero. It is spin polarization, combined with a long spin relaxation time exceeding under certain conditions one second at room temperature \cite{Maurer2012}, that opens the possibility for various applications of NV$^-$ centers.

One of the methods to study NV$^-$ centers is to research the dependence of the luminescence intensity of the NV$^-$ centers vs. applied external magnetic field (magnetic spectrum) \cite{VanOort1989,Epstein2005,Hanson2006,Rogers2008,Rogers2009,Lai2009,Armstrong2010}. Magnetic field modifies the polarization degree of NV$^-$ centers. Since luminescence intensity grows with polarization \cite{Manson2006}, a decrease of polarization due to the interaction with the external magnetic field reduces photoluminescence intensity. In addition to a smooth magnetic field dependence of luminescence intensity, sharp lines can be observed in the spectrum. However, these lines are only detectable if the magnetic field vector is parallel to axis $\langle111\rangle$ of the diamond crystal lattice. Even a slight misalignment strongly broadens the lines and suppresses their magnitude. The lines are attributed to either an anticrossing of terms in the NV$^-$ center or a resonance interaction of the NV$^-$ center with other paramagnetic defects in diamond. However, a comprehensive theoretical description of the processes giving rise to these lines still remains to be developed.

It is known that in weak magnetic fields there are effects resulting from the coherence of quantum states, e.g. the Hanle effect \cite{Hanle1924}. Another example is the zero field line in MARY (Magnetically Affected Reaction Yield) spectroscopy \cite{Woodward2010,Anisimov1983,Fischer1983}. Therefore, we specifically aimed to research the low-field part of the magnetic spectrum of NV$^-$ centers. We experimentally found a zero field line and studied its behavior, which is the subject of this work.

\section{EXPERIMENTAL}

\subsection{Samples}

The experiments were performed using five samples of an irregularly shaped synthetic diamond (see Fig.~\ref{set-up}(a)) grown at high temperature and pressure in a Fe-Ni-C system. As-grown crystals were then irradiated by electrons with energy 3~MeV. The dose was $10^{18}$~e/cm$^2$. Then the samples were annealed for 2 hours under vacuum at 800$\rm ^o$C. The state of impurity defects was monitored through the measurement of absorption spectra in the UV, VIS and IR bands. The absorption spectra in the short-wave band were measured at 300 and 77~K, in the middle IR band at 300~K. The IR absorption spectra were normalized using an internal 12.3~cm$^{-1}$ absorption standard at 1995~cm$^{-1}$ \cite{Zaitsev2001}. The single-photon part of the spectrum was found to be dominated by donor nitrogen absorption (C centers -- electrically neutral single substitutional nitrogen atoms). It is known that fast electron irradiation gives rise to vacancies in the diamond lattice which after annealing at 800$\rm ^o$C become mobile and get captured by the C centers forming NV complexes. This leads to an intense absorption in the visible band of the absorption spectra by the electron-vibration system with a zero-phonon line (ZPL) at 637~nm (1.945~eV) due to negatively charged NV$^-$ complexes. Concentration of NV$^-$ centers (in cm$^{-3}$) was estimated from the integral absorption $\mu_{637}$ at the 637~nm (in meV cm$^{-1}$) ZPL at 77~K by expression \cite{Lawson1998}:
$$N_{NV^-} = 8.9\times10^{15}\mu_{637}.$$
Concentration of the C centers was determined from the absorption at 1135~cm$^{-1}$, that of A centers (a neutral nearest-neighbor pair of nitrogen atoms substituting for the carbon atoms) at 1282~cm$^{-1}$, that of C$^+$ (positively charged single substitutional nitrogen atoms) at 1332~cm$^{-1}$, that of NE1 centers (N-V-Ni-V-N system) at 472.8~nm \cite{Nadolinny1994,Yelisseyev1996}. Measurement results for samples SL1-SL4 are shown in Table~\ref{t1}.
The fifth sample was prepared from sample SL3 by grinding it to powder with particle size within 0.1mm.

Fig.~\ref{set-up}(a) shows the photoluminescence of sample SL2 excited by blue light. The image was taken using a red optical filter. The intense red luminescence is due to negatively charged NV centers.

 \begin{table}\caption{Concentrations of various defects in the studied samples (in ppm).}

 \begin{tabular}{lcccc}

   \hline
\hline

   ~       & ~~~~~~SL1~~~~~~ & ~~~~~~SL2~~~~~~ & ~~~~~~SL3~~~~~~ & ~~~~SL4~~~~ \\

    \hline

  [NV$^-$] &  8.2    & 5.5     & 2.4     & 0.65    \\

  [C]      & 23      & 55      & 150     & $<$5    \\

  [C$^+$]  & 41      & $<$2     & 5.5     & 2.75   \\

  [A]      & 50      & $<$5     & 125     & $<$5    \\

  [NE1]    & 0.012   & $<$0.003  & $<$0.003  & $<$0.003 \\

   \hline
   \hline

  \end{tabular}
  \label{t1}
   \end{table}

\subsection{Setup}

The sample was placed in the EPR spectrometer cavity. The electric magnet of the spectrometer was equipped with special coils for reversed magnetization in order to sweep the field through zero. The transverse component of the magnetic field did not exceed 0.1~G. In our experiments, the microwave field always equaled to zero.

Samples were aligned using a manual goniometer.

\begin{figure}
   \includegraphics[width=0.4\textwidth]{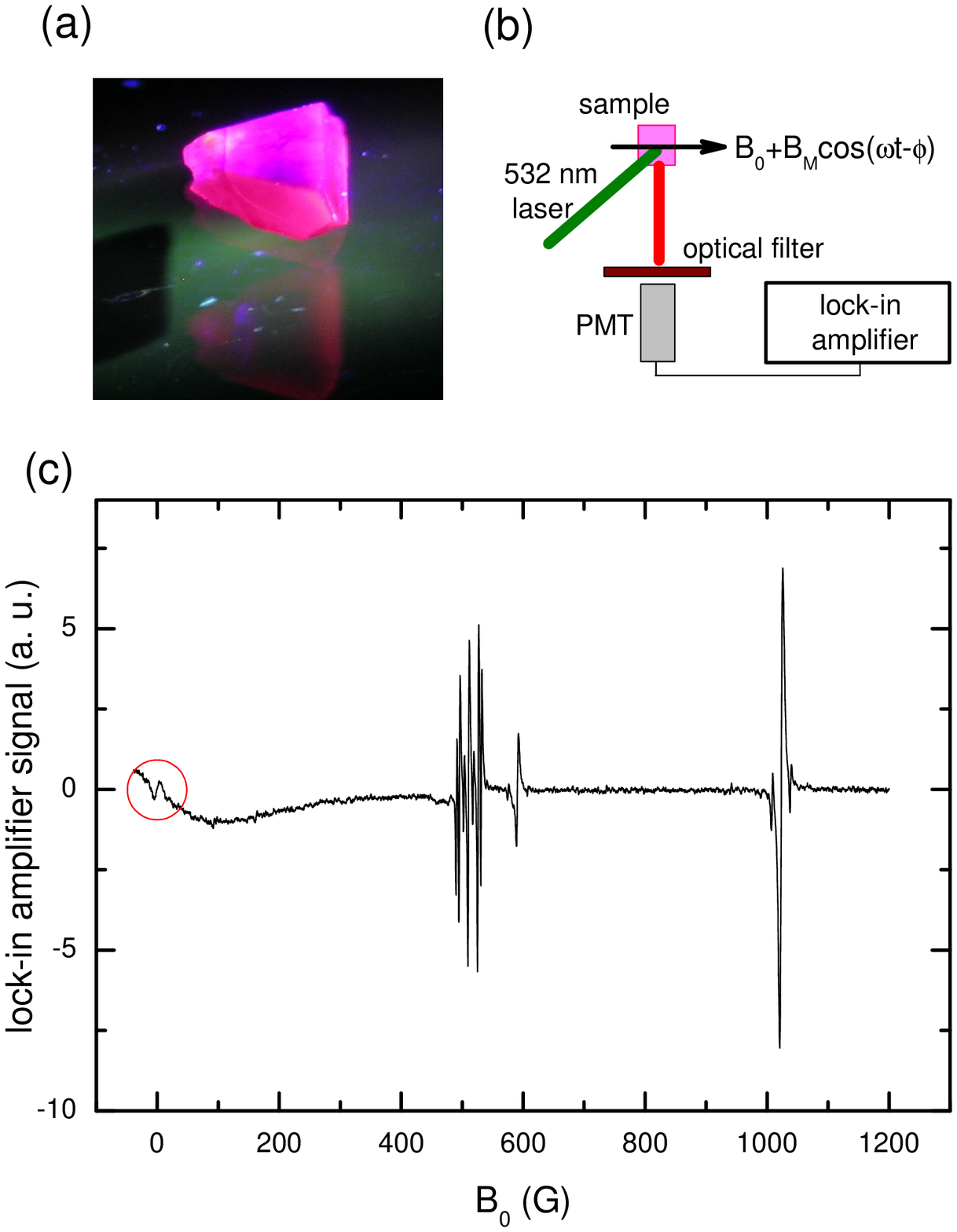} \caption{(Color online) (a) A photo of sample SL2. (b) A diagram of the setup for the observation of magnetic spectra. (c) The magnetic spectrum of sample SL2. The red circle shows the zero field line. \label{set-up}}
    \end{figure}

The sample was irradiated by a 400~mW laser beam at 532~nm. Laser power stability was within 1\%. The direction of the light beam was perpendicular to the magnetic field vector \textbf{B}$_0$ as shown in Fig.~\ref{set-up}(b). The laser beam was linearly polarized. We performed experiments with varied orientation of the polarization vector $\textbf{E}$ to the external magnetic field \textbf{B}$_0$. As demonstrated below, this variation had a considerable effect on the experimental results.

The diameter of the laser beam was within 1~mm. The width of the cavity lattice slits through which the sample was irradiated was 0.5~mm.

The light from the sample passed through a quartz lightguide and an optical filter to a photomultiplier tube (PMT). The filters were selected to reduce the light intensity down to a level acceptable for the PMT and to extract the spectral line of the negatively charged NV-center. We used a set of glass colored and neutral filters that passed the light with wavelengths longer than 680nm. The light was registered by a photomultiplier tube FEU-119 sensitive to the red spectral band.

\begin{figure}
   \includegraphics[width=0.4\textwidth]{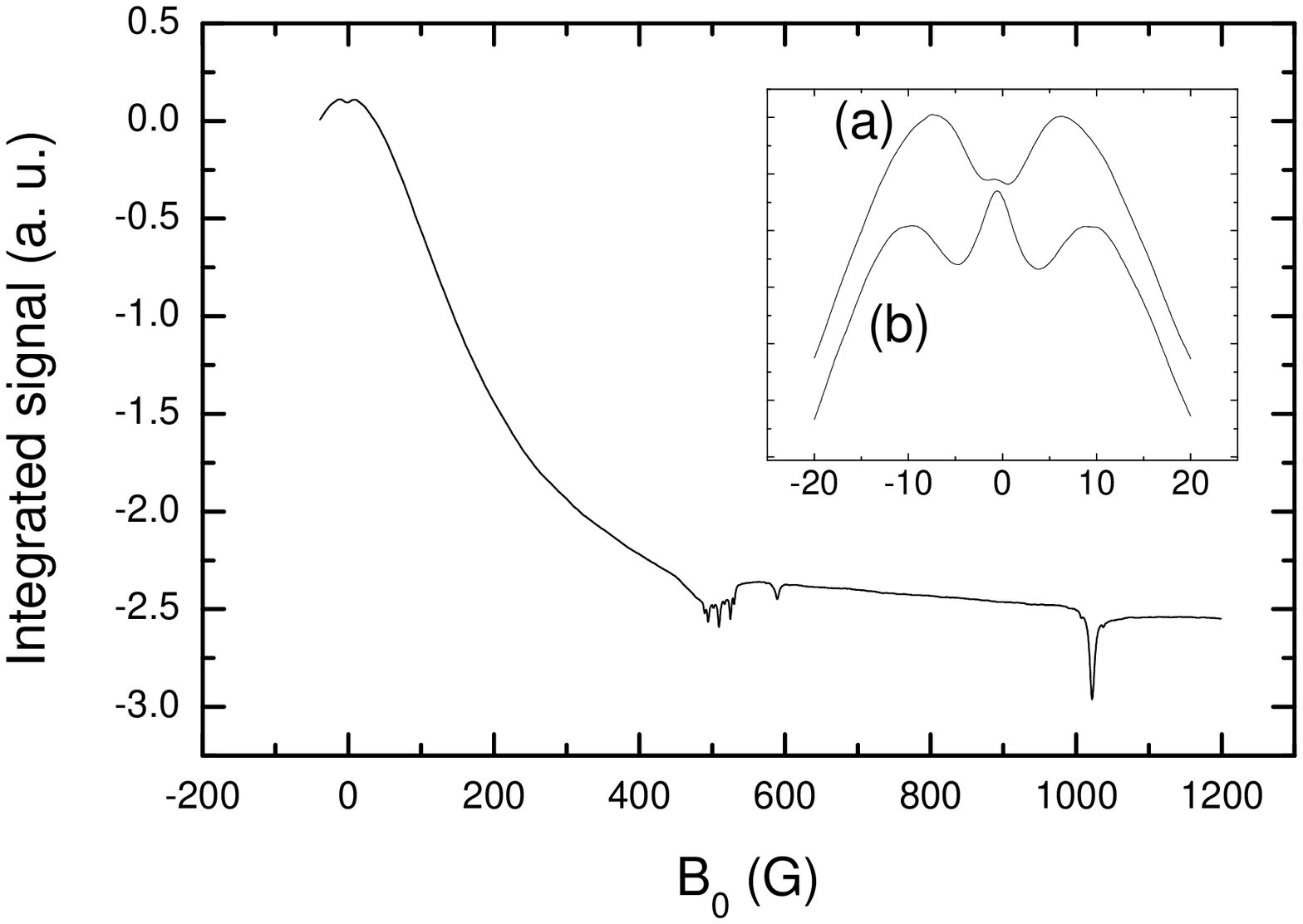} \caption{The spectrum shown in Fig.~\ref{set-up}(c) upon integration. The insert shows integrated zero field lines from Figure~\ref{1040}(a) and Fig.~\ref{1040}(b). \label{intspec}}
     \end{figure}

The signal from the PMT was fed to the input of a lock-in amplifier. The use of the lock-in amplifier considerably improved the signal-to-noise ratio in our experiments. Modulation frequency was 12.5~kHz. Modulation magnitude $\rm B_M$ for all the experimental results discussed here was 0.5~G.

The phase shift $\phi$ of the modulation magnetic field relative to the reference signal from a generator is determined by the impedance of the modulation coil and its leads as well as by the skin effect in the cavity walls. The phase shift $\theta$ of the signal coming from the sample relative to the modulation magnetic field is determined by the time function of the system response to the variation of the magnetic field. Adjusting the lock-in amplifier's phase $\phi_{LA}$ one can determine $\theta$ and thus obtain information on the time parameters of the processes going in the sample. For an exponential system response, the characteristic time $\tau$ of a process is related to the phase shift $\theta$ by the following expression: $\tan \theta = \omega \tau$, where $\omega$ is modulation frequency.

We determined the value of $\phi$ from experiments with recombination fluorescence in X-irradiated nonpolar liquids because in this case the processes determining the system response to the variation of the magnetic field occur in the nanosecond time domain. Consequently, $\tau \ll 2\pi/\omega$, and the signal from the lock-in amplifier reaches its maximum when $\phi_{LA}=\phi$. As demonstrated below, when measuring magnetic spectra of NV$^-$ centers one has to select $\phi_{LA}=\phi+\theta_{LA}$ considerably different from $\phi$. In all the figures in this paper we show experimental results for $\theta_{LA}=70{\rm ^o}$.

All the experiments were carried out at room temperature. An aluminum radiator was used to dissipate the heat from the sample.

\section{RESULTS}

Fig.~\ref{set-up}(c) shows an experimentally registered dependence of the lock-in amplifier output signal vs. external magnetic field for sample SL2. Hereinafter this dependence is referred to as magnetic spectrum. The lock-in amplifier signal can be considered proportional to the derivative of the sample luminescence intensity vs. magnetic field.
Fig.~\ref{intspec} shows the integrated spectrum of sample SL2 corresponding to the curve shown in Fig.~\ref{set-up}(c). The integrated spectrum looks similar to previously observed \cite{Rogers2008,Rogers2009} experimental curves.

As one can see in Fig.~\ref{set-up}(c), the spectrum consists of a wide line with a maximum (in the absolute units) around 100~G and many narrow lines. The wide line describes the decay of luminescence intensity due to a decrease of the polarization degree of the NV$^-$ center driven by the magnetic field component perpendicular to the center's symmetry axis. This effect is not observed in NV$^-$ centers where the symmetry axis is parallel to the axis of the external magnetic field \textbf{B}$_0$. However, this orientation activates other polarization-reducing mechanisms giving rise to narrow lines in the magnetic spectrum clearly seen in Fig.~\ref{set-up}(c). Most of these lines were observed before \cite{VanOort1989,Epstein2005,Hanson2006,Rogers2008,Rogers2009,Armstrong2010}. The physical backgrounds of these lines are different.

The most intense line at $\sim$1028~G is due to an level anticrossing (LAC) of the triplet terms of the ground state of the NV$^-$ center with different spin projections on the symmetry axis. The nature of the other lines has not been established with certainty. They are usually referred to as cross-relaxation lines \cite{VanOort1989}. The line in the field $\sim$600~G is reasonably attributed to the counteraction between NV$^-$ centers with different orientations \cite{VanOort1989,Armstrong2010}, while most of the lines in the 490--540~G range are attributed to the interaction of the NV$^-$ centers with electrically neutral single substitutional nitrogen atoms \cite{VanOort1989,Hanson2006,Armstrong2010}.

Registering a magnetic spectrum with resolved narrow lines would require a very fine alignment of the crystal so that its $\langle111\rangle$ axis would be parallel to the external magnetic field. A misalignment from this orientation strongly broadens the lines and reduces their magnitude.
One of the indicators of the correct alignment of the crystal is the shape of the line at $\sim$600~G (in our experiments the line was centered at 590~G). A slight misalignment of the crystal $\langle111\rangle$ axis to the external magnetic field splits this line to three lines of approximately equal magnitude \cite{Armstrong2010}. However, as our experiments showed, when the crystal is precisely aligned, this line may have satellites of lower magnitude (see Fig.~\ref{500}). Another measure of crystal alignment precision is the presence of satellites at the LAC line which disappear when the misalignment is as little as 0.2$\rm ^o$  \cite{Armstrong2010}.

The red circle in Fig.~\ref{set-up}(c) marks a narrow line in zero field never observed before. This research is focused on the properties of this line.

\begin{figure}
   \includegraphics[width=0.4\textwidth]{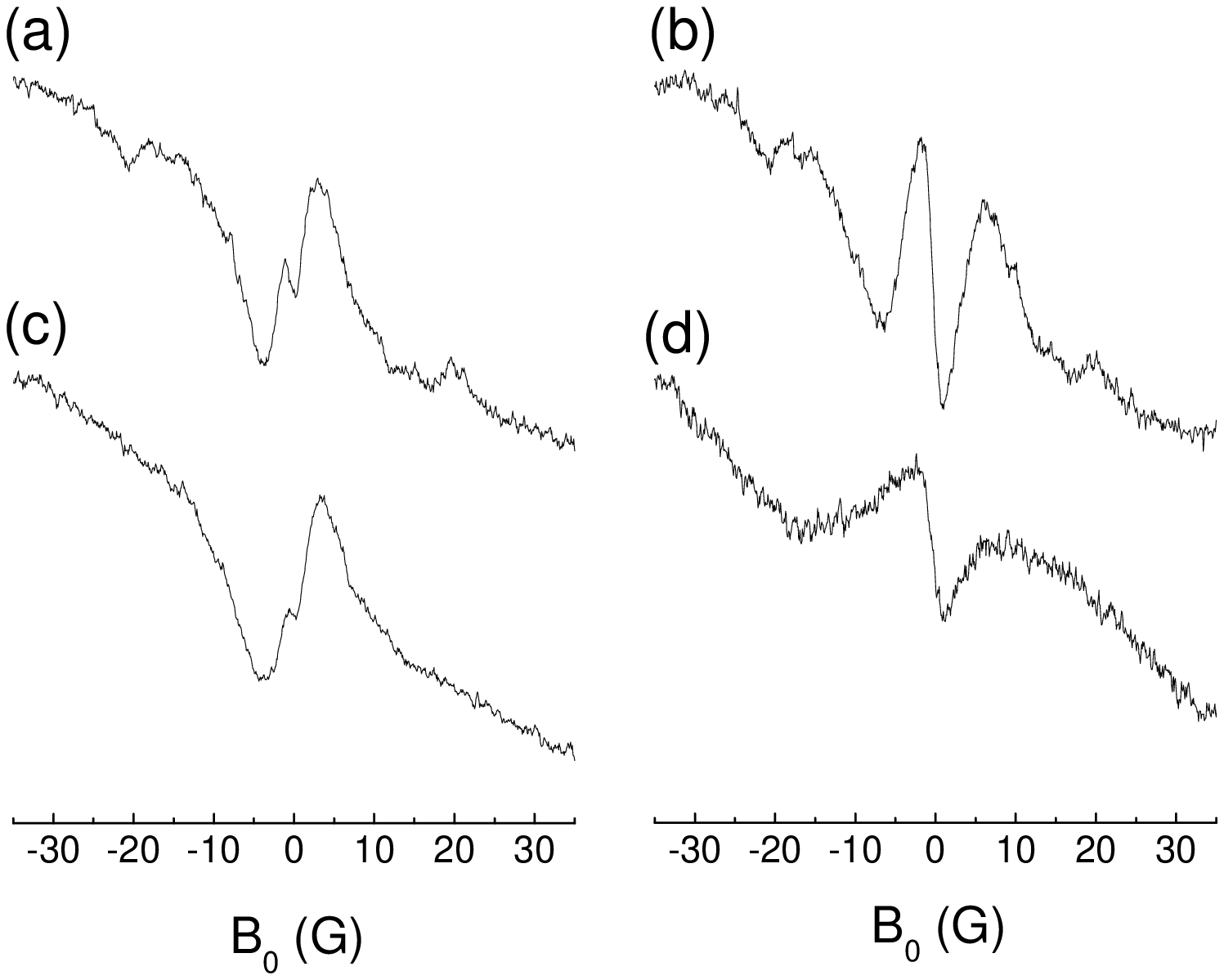} \caption{Zero field line in sample SL1. The $\langle111\rangle$ axis is oriented along the magnetic field \textbf{B}$_0$ ((a) and (b)) and rotated by 90$\rm ^o$ relative to \textbf{B}$_0$ ((c) and (d)). The polarization vector of the excitation light \textbf{E} is parallel ((a) and (c)) or perpendicular ((b) and (d)) to magnetic field \textbf{B}$_0$. \label{1040}}
    \end{figure}

Fig.~\ref{1040} shows magnetic spectra of sample SL1 in low fields for various orientations of the sample and various orientations of the external magnetic field \textbf{B}$_0$ to the polarization vector \textbf{E} of the excitation laser beam.

Curves (a) and (b) correspond to crystal orientation \textbf{B}$_0~||~\langle111\rangle$. The crystal was aligned manually. The criterion was the presence of the narrow LAC line at $\sim$1028~G with resolved satellite lines and a single narrow line at $\sim$600~G (also having satellites of a lower magnitude in this sample). This observation indicated that alignment precision was 0.2$\rm ^o$ or better. The position of the other axes was not controlled. Crystal orientation in cases (a) and (b) was the same.

Curves (c) and (d) correspond to crystal orientation \textbf{B}$_0~\perp~\langle111\rangle$. This orientation was achieved the following way: first the condition \textbf{B}$_0~||~\langle111\rangle$ was ensured as described above and then the crystal was rotated by 90$\rm ^o$ around an axis perpendicular to \textbf{B}$_0$ randomly oriented to the crystal. Therefore, crystal orientation in this case is undetermined to a large extent. For curves (c) and (d) crystal orientation was different as they were obtained in different experimental series.

\begin{figure}
   \includegraphics[width=0.4\textwidth]{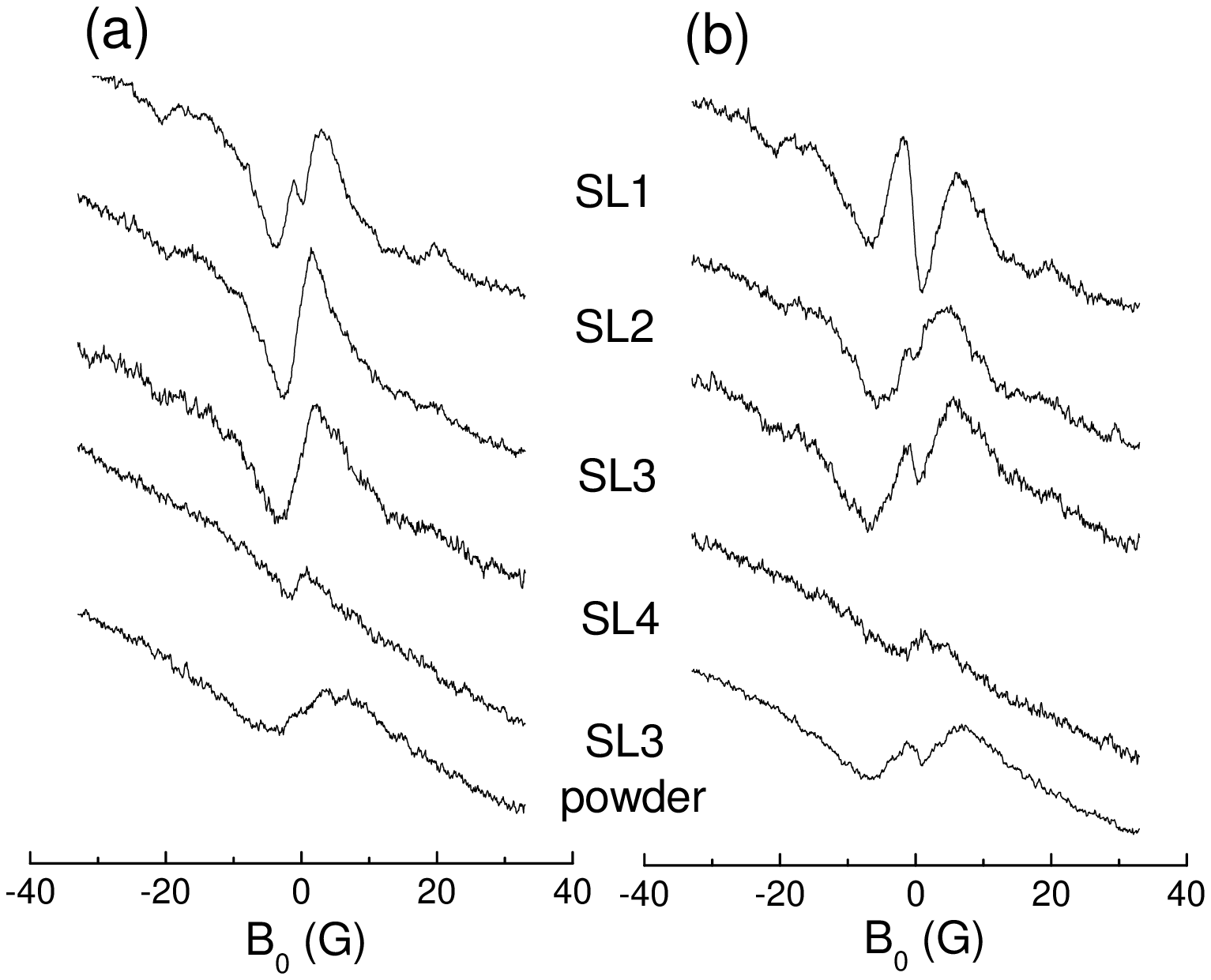} \caption{Zero field line in single crystals SL1--SL4 and in a powdered sample prepared from SL3. The $\langle111\rangle$ axis in all the single crystals is oriented along the magnetic field \textbf{B}$_0$. The polarization vector of the excitation light \textbf{E} is parallel (a) or perpendicular (b) to the magnetic field \textbf{B}$_0$. \label{all_s}}
    \end{figure}

For curves (a) and (c) the polarization vector \textbf{E} of the laser beam is parallel to the magnetic field vector \textbf{B}$_0$ while for curves (c) and (d) they are perpendicular. Signal magnitude for all the curves was normalized to sample luminescence intensity in zero field.

Curves (a) and (b) upon integration are shown in the insert in Fig.~\ref{intspec}. The lock-in measurement technique cannot determine the absolute magnitude of the effect as it registers its first derivative. Although, having performed additional experiments to measure the luminescence intensity, we found that the variation of the zero field line intensity was within 1 per cent.

As one can see in Fig.~\ref{1040}, in all the cases a zero field line is observed with a magnitude varying to a small extent when the sample is rotated or the polarization of the excitation light is varied. For \textbf{E}$\perp$\textbf{B}$_0$ a narrow inverted line appears in zero field. Its width (peak-to-peak) is about 2~G.

As one can see in the Figure, a weak inverted line in zero field is also observed when the polarization vector of the excitation light is parallel to the field \textbf{B}$_0$. This is apparently due to the fact that we can only control light polarization \emph{outside} the sample. Because the light is refracted when it enters the sample, its propagation direction can change giving rise to a non-zero perpendicular component of light polarization \emph{inside} the sample. It is this component that accounts for the inverted line. As one can see in Fig.~\ref{all_s}, there was no such an effect in the other samples.

A substantial feature of the spectrum is the presence of a satellite line just below 20~Gs. The magnitude of this line is independent of excitation light polarization. However, it disappears when the sample is rotated, as shown in Fig.~\ref{1040}(c). In Fig.~\ref{1040}(d) this line appears in a very much broadened form.

Fig.~\ref{all_s} shows magnetic spectra of different samples for two different polarizations of the excitation light. All the spectra are normalized to the sample luminescence intensity in zero magnetic field. In the monocrystalline samples SL1-Sl4 the concentration of NV$^-$ centers monotonously decayed. The bottom spectrum corresponds to a powdered sample prepared by grinding the single-crystal sample SL3. For all the samples (except the powdered one) \textbf{B}$_0~||~\langle111\rangle$.

As one can see in the Figure, in all the samples there is a zero field line. In samples SL1--Sl3 the line's magnitude is the same while in sample SL4 (with the lowest concentration of the NV$^-$ centers) it sharply drops and the width of the line becomes much smaller. At the same time, in all the samples an inverted narrow line appears in zero when the polarization of the excitation light changes from parallel toward perpendicular to the external magnetic field \textbf{B}$_0$. When the polarization is parallel the narrow inverted zero field line is absent in all the samples except SL1. In samples SL1--Sl3 a satellite line at 20~G is also observed; its magnitude is independent of excitation light polarization and decays when the concentration of the NV$^-$ centers decreases.

\begin{figure}
   \includegraphics[width=0.4\textwidth]{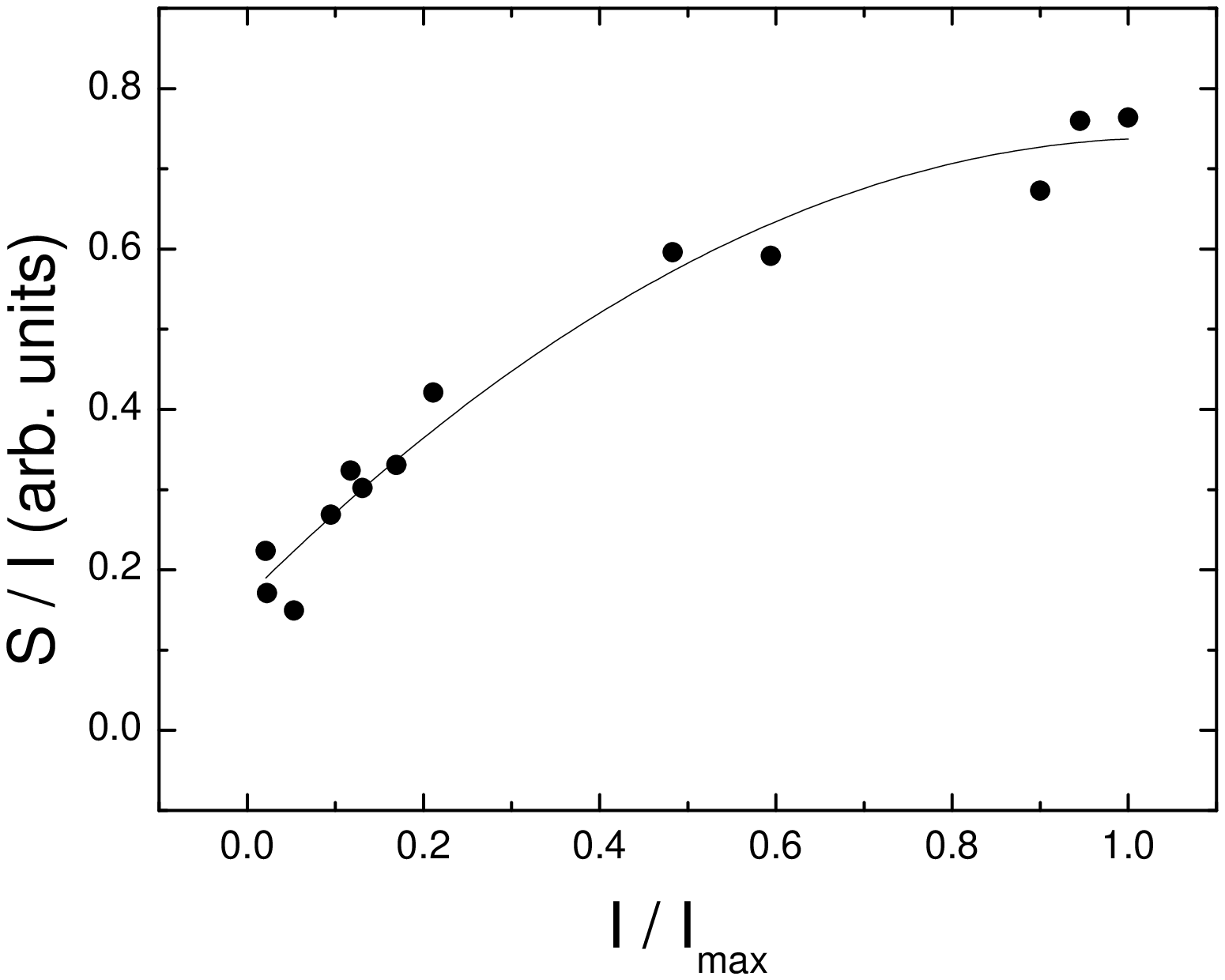} \caption{The dependence of the normalized signal magnitude vs. relative sample luminescence intensity (closed circles). The solid line shows a parabolic approximation of the experimental data.
      \label{intens_dep}}
    \end{figure}

Fig.~\ref{intens_dep} shows the dependence of the magnitude of the zero field line normalized to the sample light intensity vs. the relative spectrum light intensity. In the experiment, the SL1 sample was oriented in a way that the $\langle111\rangle$ axis of the crystal was perpendicular to magnetic field \textbf{B}$_0$. The polarization vector of the excitation laser beam was parallel to \textbf{B}$_0$. The laser beam was attenuated by optical filters. A magnetic spectrum measured without attenuation filters is shown in Fig.~\ref{1040}(c).

The magnitude $S$ of the line was calculated as follows: from the zero field line spectrum a straight line was subtracted to compensate for the slope, then the vertical peak-to-peak distance was measured. The magnitude was normalized by the division by the photomultiplier tube current $I$ in zero field which was proportional to the sample luminescence intensity at that field. The X-axis shows the ratio of the current with the optical filter used to the current with no filter $I/I_{max}$, which is proportional to the relative sample luminescence intensity in zero magnetic field.

The scattering of the points is due to the fact that different filters not only attenuate the light but also disperse it differently. The magnitude of the zero field line depends on sample illumination while the overall luminescence intensity depends on the illumination and size of the light spot that we could not control. Therefore the measured sample light intensity is not a perfect value.

The solid line in the Figure shows the best parabolic fit of the experimental data $S/I=a+b(I/I_{max})+c(I/I_{max})^2$, where $S$ is the magnitude of the zero field line, $I$ is the PMT current at B$_0$=0 proportional to the sample light intensity, $a=0.017$, $b=0.109$, $c=-0.052$. From our experiments we can conclude that the magnitude of the zero field line has a non-linear relation to the intensity of the excitation light in the sample. In the low-intensity region this dependence is close to a square law.

\section{MODEL}

Our experiments suggest that the properties of the zero field line are considerably different from those of any other lines previously known for the magnetic spectrum of the NV$^-$ centers in diamond. Why this line appears cannot be accounted for by the processes in an isolated center. For example, in the theoretical calculations by Rogers et al. \cite{Rogers2009} there is no such line.

We also considered an exotic model of geminate pairs in diamond under two-photon photoionization of the NV$^-$ center forming a neutral NV$^0$center and a free electron similar to those discussed by Siyushev et al. \cite{Siyushev2013}. If this electron is captured by a nearby positively charged nitrogen atom N$^+$ with further spin evolution in the geminate pair N$^0$--NV$^0$ followed by a spin-selective electron transfer to the NV$^0$ center then the population of the triplet states of thus formed NV$^-$ center with different spin projections on the center's symmetry axis is magnetic field-dependent. Consequently, in this case the luminescence intensity will depend on the magnetic field and give a low-field line in the spectrum. This model accounts for the square-law dependence of the line's magnitude vs. excitation light intensity. However, calculations within this model fail to reproduce the properties of the line observed in the experiment. For example, the predicted lines are inverted when crystal orientation is changed, in contrast to the experimental observation. Also, the probability of these processes is too low to give a marked line magnitude.

We suppose that the zero field line in the magnetic spectrum can be explained by the dipole-dipole interaction between the NV$^-$ centers. A similar model explaining the nature of the line at $\sim$600~G was suggested in \cite{Armstrong2010}.

The ground state of the NV$^-$ center is triplet, so the center is a magnetic dipole. The magnetic dipoles of two centers interact, which can lead to transitions between the triplet sub-levels in each of the NV$^-$ centers. For this interaction to affect the luminescence quantum yield, it has to modify the overall population of both triplet states with zero projections on the symmetry axes of both centers in the interacting pair. Obviously, if the orientation of the centers is the same the overall population will not change. However, if the orientation is different then the dipole-dipole interaction can modify this overall population of the centers under different degree of their polarization.

The efficiency of such a modification of the population of the terms in a pair of NV$^-$ centers considerably depends on the external magnetic field. This is due to the fact that in zero field the transitions between the energy terms of differently-oriented centers have a resonance nature but when an external magnetic field is applied the terms in such centers shift differently and go out of resonance. This slows down the mixing processes and even stops them in higher fields. This should lead to a drop in the zero of the luminescence intensity vs. external magnetic field curve.

A different degree of the polarization of differently-oriented NV$^-$ centers is achieved under sample excitation by polarized light. This is due to the $C_{3v}$-symmetry of the NV$^-$ center. The transition from the ground state $^3A_2$ to an exited state $^3E$ is forbidden when the polarization vector of the excitation light \textbf{E} is parallel to the symmetry axis. Therefore the luminescence intensity and the degree of polarization of the center depend on the direction of vector \textbf{E} in the excitation light and grow with the transverse component of \textbf{E}.

\begin{figure}[b]
   \includegraphics[width=0.4\textwidth]{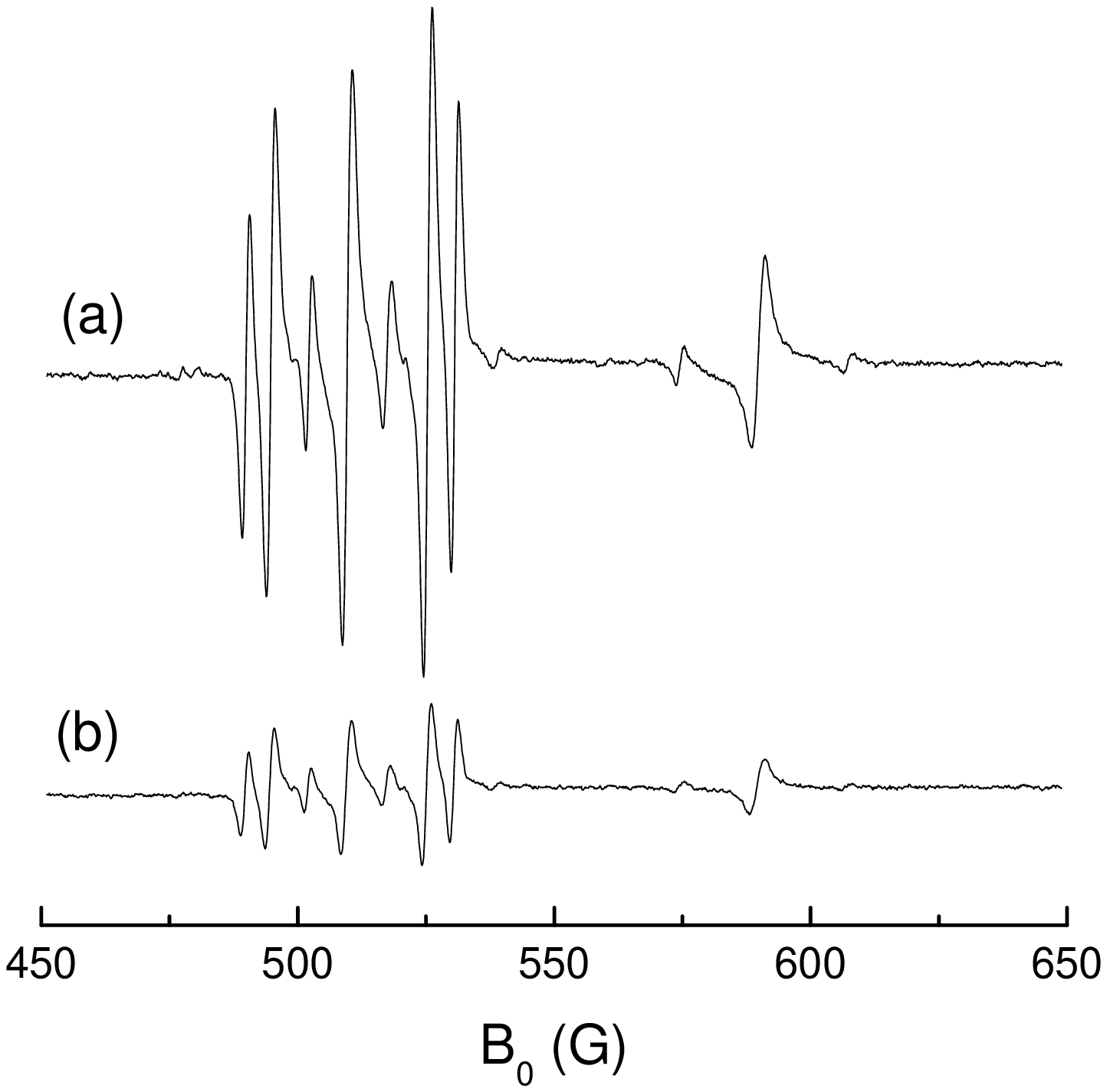} \caption{The magnetic spectrum of sample SL2 for $\langle111\rangle~||~$\textbf{B}$_0$. The polarization vector of the excitation light \textbf{E} is perpendicular to the magnetic field \textbf{B}$_0$ (a) or parallel to \textbf{B}$_0$ (b). The spectra are normalized to sample luminescence intensity in zero magnetic field. \label{500}}
    \end{figure}

In order to test the outcome of this effect we performed an experiment aimed to observe the effect of the variation of the excitation light polarization on the narrow lines in the magnetic spectrum of NV$^-$ centers. Fig.~\ref{500} shows the magnetic spectrum of sample SL2 for \textbf{B}$_0~||~\langle111\rangle$ under various polarizations of the excitation light. The figure shows a spectrum segment between 450 to 650~G. In Fig.~\ref{500}(a) the polarization vector is perpendicular to \textbf{B}$_0$ while in Fig.~\ref{500}(b) they are parallel. The spectra are normalized to sample luminescence intensity in zero magnetic field. As one can see in the figure, the spectra are almost identical and only differ in magnitudes, which are different approximately by a factor of 4. We should mention that this difference is in a marked contrast to the transformation of the spectrum in low fields where the shape of the spectrum changes greatly while its magnitude alters only slightly. Since the narrow lines in the spectrum shown in Fig.~\ref{500} are only generated by NV$^-$ centers with symmetry axes oriented parallel to the external magnetic field we can conclude that in our experiment, a change in the orientation between the magnetic field and the light polarization vector should also suppress the efficiency of the excitation of the centers by a factor of 4.

The results of this experiment suggest that under polarized light-excitation, NV$^-$ centers differently oriented in the crystal can have considerably different ground triplet state polarizations. This means that the difference between the populations of the terms for different spin projections on the center's symmetry axis is considerably different. Therefore the dipole-dipole interaction between the electron spins considerably modifies the overall population of the terms with zero spin projection. Consequently, the luminescence yield in low magnetic fields is modified.

The change in sample luminescence intensity due to the dipole-dipole interaction between NV$^-$ centers is proportional to the excitation light intensity multiplied by the difference between the polarization degrees of the centers with different spatial orientation. The difference between the polarization degrees, in its turn, is proportional to excitation light intensity, for moderate intensities. Therefore, the magnitude of the observed effect vs. excitation light intensity should follow a square law.

As was mentioned above, the zero field line has a considerable phase shift. This phase shift cannot be accurately determined as it is slightly different for different segments of the line. As a consequence, the line cannot be zeroed by the adjustment of the lock-in amplifier phase $\theta_{LA}$. For sample SL1 the phase shift is $\theta=80\pm2^{\rm o}$. One can estimate the exponential response time as $\tau=\tan \theta / \omega \approx 7\times 10^{-5}$~s.

Suppose the response time $\tau$ is determined by the energy of the dipole-dipole interaction between two NV$^-$ centers which can be crudely estimated by a dimension factor $V=(g \mu_B)^2/(h r^3)$, where $g$ is free electron g-value, $\mu_B$ is Bohr's magneton, $h$ is Planck's constant, $r$ is separation between the centers. Assuming $V=1/\tau$ we get $r\approx15$~nm, which is of the same order of magnitude as the average separation between the centers in the sample.

On the other hand, $\tau$ is of the same order of magnitude as the typical phase spin relaxation times in such samples \cite{Kennedy2003}. Therefore it cannot be excluded that it is phase relaxation time that determines the phase shift.

In order to test our hypothesis we performed experiments using an incoherent non-polarized light source to excite the sample. As a source of this kind we used a 10~W green LED. The LED light was focused on the sample by a lens. In this experiment we successfully observed both the wide line with a maximum at 100~G and narrow lines in the magnetic spectrum above 400~G. There was no zero field line at all. This experiment supports our hypothesis although cannot be taken as its final proof because we might fail to adequately focus the LED light on the sample to achieve an illumination similar to that given by the laser. Although, the sample luminescence intensity was comparable to that observed under laser irradiation with the attenuation filters engaged when we detected the zero field line.

We have not found a straightforward explanation to account for the transformation of the spectrum when the excitation light polarization vector changes as well as for the rise of the satellite lines at about 20~G. As for the satellite lines, their nature might be similar to that of the satellites to the line at $\sim$600~G. The splitting values and magnitude ratios relative to the main line are very close in both cases. Moreover, it is possible that two drastic decreases of the relaxation time at 0 and $\sim$600~G in relaxation time vs. magnetic field dependence \cite{VanOort1989} can be explained equally in terms of dipole-dipole interaction between two NV$^-$ centers.

In order to explain the behavior of the discovered line within the hypothesis we suggest, the experiment has to be compared to a numerical model. Preliminary calculations we made within the model of the dipole-dipole interaction between two NV$^-$ centers have predicted that this zero field line should appear. Also, for a certain set of parameters a narrow inverted line is predicted to appear in the center of the zero field line.

\section{CONCLUSION}

We have found a zero field line in the magnetic spectrum of diamonds containing NV$^-$ centers. This line has the following properties:

   1) The magnitude of the line weakly depends on sample orientation.

   2) The shape of the line strongly depends on excitation light polarization. If the polarization is perpendicular to the external magnetic field, an inverted line in zero field appears.

   3) When the concentration of the NV$^-$ centers in the sample strongly decreases the magnitude of the line drops. At the same time, the width of the line decreases.

   4) At higher concentrations of the NV$^-$ centers, additional lines in the spectrum appear at about 20~G. This line is only observed for \textbf{B}$_0~||~\langle111\rangle$. Its magnitude is independent of excitation light polarization.

   5) The magnitude of the zero field line reaches its maximum at a large (about 80$\rm ^o$) phase shift $\theta_{LA}$ of the lock-in amplifier suggesting that the processes accounting for the rise of this line are quite slow (microsecond time domain).

   6) The line is also observed in a powdered sample where its magnitude is comparable to that of the line in the single crystal.

   7) The magnitude of the line strongly depends on excitation light intensity. Besides, for low intensities this dependence is close to a square law.

We suggest that this line appears due to the magnetic dipole-dipole interaction between NV$^-$ centers differently oriented in the crystal that modifies the polarization degree of the ground triplet state.

\begin{acknowledgments}
Authors are grateful to Yu.N.~Molin and A.A.~Shubin for helpful discussions. The work was supported by the Program `Leading Scientific Schools' (project No NSh 5744.2014.3).
\end{acknowledgments}

\bibliography{Anishchik}

\begin{thebibliography}{36}%
\makeatletter
\providecommand \@ifxundefined [1]{%
 \@ifx{#1\undefined}
}%
\providecommand \@ifnum [1]{%
 \ifnum #1\expandafter \@firstoftwo
 \else \expandafter \@secondoftwo
 \fi
}%
\providecommand \@ifx [1]{%
 \ifx #1\expandafter \@firstoftwo
 \else \expandafter \@secondoftwo
 \fi
}%
\providecommand \natexlab [1]{#1}%
\providecommand \enquote  [1]{``#1''}%
\providecommand \bibnamefont  [1]{#1}%
\providecommand \bibfnamefont [1]{#1}%
\providecommand \citenamefont [1]{#1}%
\providecommand \href@noop [0]{\@secondoftwo}%
\providecommand \href [0]{\begingroup \@sanitize@url \@href}%
\providecommand \@href[1]{\@@startlink{#1}\@@href}%
\providecommand \@@href[1]{\endgroup#1\@@endlink}%
\providecommand \@sanitize@url [0]{\catcode `\\12\catcode `\$12\catcode
  `\&12\catcode `\#12\catcode `\^12\catcode `\_12\catcode `\%12\relax}%
\providecommand \@@startlink[1]{}%
\providecommand \@@endlink[0]{}%
\providecommand \url  [0]{\begingroup\@sanitize@url \@url }%
\providecommand \@url [1]{\endgroup\@href {#1}{\urlprefix }}%
\providecommand \urlprefix  [0]{URL }%
\providecommand \Eprint [0]{\href }%
\providecommand \doibase [0]{http://dx.doi.org/}%
\providecommand \selectlanguage [0]{\@gobble}%
\providecommand \bibinfo  [0]{\@secondoftwo}%
\providecommand \bibfield  [0]{\@secondoftwo}%
\providecommand \translation [1]{[#1]}%
\providecommand \BibitemOpen [0]{}%
\providecommand \bibitemStop [0]{}%
\providecommand \bibitemNoStop [0]{.\EOS\space}%
\providecommand \EOS [0]{\spacefactor3000\relax}%
\providecommand \BibitemShut  [1]{\csname bibitem#1\endcsname}%
\let\auto@bib@innerbib\@empty
\bibitem [{\citenamefont {Doherty}\ \emph {et~al.}(2013)\citenamefont
  {Doherty}, \citenamefont {Manson}, \citenamefont {Delaney}, \citenamefont
  {Jelezko}, \citenamefont {Wrachtrup},\ and\ \citenamefont
  {Hollenberg}}]{Doherty2013}%
  \BibitemOpen
  \bibfield  {author} {\bibinfo {author} {\bibfnamefont {M.~W.}\ \bibnamefont
  {Doherty}}, \bibinfo {author} {\bibfnamefont {N.~B.}\ \bibnamefont {Manson}},
  \bibinfo {author} {\bibfnamefont {P.}~\bibnamefont {Delaney}}, \bibinfo
  {author} {\bibfnamefont {F.}~\bibnamefont {Jelezko}}, \bibinfo {author}
  {\bibfnamefont {J.}~\bibnamefont {Wrachtrup}}, \ and\ \bibinfo {author}
  {\bibfnamefont {L.~C.~L.}\ \bibnamefont {Hollenberg}},\ }\href {\doibase
  10.1016/j.physrep.2013.02.001} {\bibfield  {journal} {\bibinfo  {journal}
  {Physics Reports}\ }\textbf {\bibinfo {volume} {528}},\ \bibinfo {pages} {1}
  (\bibinfo {year} {2013})}\BibitemShut {NoStop}%
\bibitem [{\citenamefont {Gruber}\ \emph {et~al.}(1997)\citenamefont {Gruber},
  \citenamefont {Dr\"{a}benstedt}, \citenamefont {Tietz}, \citenamefont
  {Fleury}, \citenamefont {Wrachtrup},\ and\ \citenamefont {von
  Borczyskowski}}]{Gruber1997}%
  \BibitemOpen
  \bibfield  {author} {\bibinfo {author} {\bibfnamefont {A.}~\bibnamefont
  {Gruber}}, \bibinfo {author} {\bibfnamefont {A.}~\bibnamefont
  {Dr\"{a}benstedt}}, \bibinfo {author} {\bibfnamefont {C.}~\bibnamefont
  {Tietz}}, \bibinfo {author} {\bibfnamefont {L.}~\bibnamefont {Fleury}},
  \bibinfo {author} {\bibfnamefont {J.}~\bibnamefont {Wrachtrup}}, \ and\
  \bibinfo {author} {\bibfnamefont {C.}~\bibnamefont {von Borczyskowski}},\
  }\href {\doibase 10.1126/science.276.5321.2012} {\bibfield  {journal}
  {\bibinfo  {journal} {Science}\ }\textbf {\bibinfo {volume} {276}},\ \bibinfo
  {pages} {2012} (\bibinfo {year} {1997})}\BibitemShut {NoStop}%
\bibitem [{\citenamefont {Wrachtrup}\ \emph {et~al.}(2001)\citenamefont
  {Wrachtrup}, \citenamefont {Kilin},\ and\ \citenamefont
  {Nizovtsev}}]{Wrachtrup2001}%
  \BibitemOpen
  \bibfield  {author} {\bibinfo {author} {\bibfnamefont {J.}~\bibnamefont
  {Wrachtrup}}, \bibinfo {author} {\bibfnamefont {S.~Y.}\ \bibnamefont
  {Kilin}}, \ and\ \bibinfo {author} {\bibfnamefont {A.~P.}\ \bibnamefont
  {Nizovtsev}},\ }\href@noop {} {\bibfield  {journal} {\bibinfo  {journal}
  {Optics and Spectroscopy}\ }\textbf {\bibinfo {volume} {91}},\ \bibinfo
  {pages} {429} (\bibinfo {year} {2001})}\BibitemShut {NoStop}%
\bibitem [{\citenamefont {Jelezko}\ \emph {et~al.}(2004)\citenamefont
  {Jelezko}, \citenamefont {Gaebel}, \citenamefont {Popa}, \citenamefont
  {Domhan}, \citenamefont {Gruber},\ and\ \citenamefont
  {Wrachtrup}}]{Jelezko2004}%
  \BibitemOpen
  \bibfield  {author} {\bibinfo {author} {\bibfnamefont {F.}~\bibnamefont
  {Jelezko}}, \bibinfo {author} {\bibfnamefont {T.}~\bibnamefont {Gaebel}},
  \bibinfo {author} {\bibfnamefont {I.}~\bibnamefont {Popa}}, \bibinfo {author}
  {\bibfnamefont {M.}~\bibnamefont {Domhan}}, \bibinfo {author} {\bibfnamefont
  {A.}~\bibnamefont {Gruber}}, \ and\ \bibinfo {author} {\bibfnamefont
  {J.}~\bibnamefont {Wrachtrup}},\ }\href {\doibase
  10.1103/PhysRevLett.93.130501} {\bibfield  {journal} {\bibinfo  {journal}
  {Phys. Rev. Lett.}\ }\textbf {\bibinfo {volume} {93}},\ \bibinfo {pages}
  {130501} (\bibinfo {year} {2004})}\BibitemShut {NoStop}%
\bibitem [{\citenamefont {Childress}\ \emph {et~al.}(2006)\citenamefont
  {Childress}, \citenamefont {{Gurudev Dutt}}, \citenamefont {Taylor},
  \citenamefont {Zibrov}, \citenamefont {Jelezko}, \citenamefont {Wrachtrup},
  \citenamefont {Hemmer},\ and\ \citenamefont {Lukin}}]{Childress2006}%
  \BibitemOpen
  \bibfield  {author} {\bibinfo {author} {\bibfnamefont {L.}~\bibnamefont
  {Childress}}, \bibinfo {author} {\bibfnamefont {M.~V.}\ \bibnamefont
  {{Gurudev Dutt}}}, \bibinfo {author} {\bibfnamefont {J.~M.}\ \bibnamefont
  {Taylor}}, \bibinfo {author} {\bibfnamefont {A.~S.}\ \bibnamefont {Zibrov}},
  \bibinfo {author} {\bibfnamefont {F.}~\bibnamefont {Jelezko}}, \bibinfo
  {author} {\bibfnamefont {J.}~\bibnamefont {Wrachtrup}}, \bibinfo {author}
  {\bibfnamefont {P.~R.}\ \bibnamefont {Hemmer}}, \ and\ \bibinfo {author}
  {\bibfnamefont {M.~D.}\ \bibnamefont {Lukin}},\ }\href {\doibase
  10.1126/science.1131871} {\bibfield  {journal} {\bibinfo  {journal}
  {Science}\ }\textbf {\bibinfo {volume} {314}},\ \bibinfo {pages} {281}
  (\bibinfo {year} {2006})}\BibitemShut {NoStop}%
\bibitem [{\citenamefont {Wrachtrup}\ and\ \citenamefont
  {Jelezko}(2006)}]{Wrachtrup2006}%
  \BibitemOpen
  \bibfield  {author} {\bibinfo {author} {\bibfnamefont {J.}~\bibnamefont
  {Wrachtrup}}\ and\ \bibinfo {author} {\bibfnamefont {F.}~\bibnamefont
  {Jelezko}},\ }\href {\doibase 10.1088/0953-8984/18/21/S08} {\bibfield
  {journal} {\bibinfo  {journal} {J. Phys.: Condens. Matter}\ }\textbf
  {\bibinfo {volume} {18}},\ \bibinfo {pages} {S807} (\bibinfo {year}
  {2006})}\BibitemShut {NoStop}%
\bibitem [{\citenamefont {Hanson}\ \emph
  {et~al.}(2006{\natexlab{a}})\citenamefont {Hanson}, \citenamefont {Gywat},\
  and\ \citenamefont {Awschalom}}]{Hanson2006b}%
  \BibitemOpen
  \bibfield  {author} {\bibinfo {author} {\bibfnamefont {R.}~\bibnamefont
  {Hanson}}, \bibinfo {author} {\bibfnamefont {O.}~\bibnamefont {Gywat}}, \
  and\ \bibinfo {author} {\bibfnamefont {D.~D.}\ \bibnamefont {Awschalom}},\
  }\href {\doibase 10.1103/PhysRevB.74.161203} {\bibfield  {journal} {\bibinfo
  {journal} {Phys. Rev. B}\ }\textbf {\bibinfo {volume} {74}},\ \bibinfo
  {pages} {161203} (\bibinfo {year} {2006}{\natexlab{a}})}\BibitemShut
  {NoStop}%
\bibitem [{\citenamefont {Gaebel}\ \emph {et~al.}(2006)\citenamefont {Gaebel},
  \citenamefont {Domhan}, \citenamefont {Popa}, \citenamefont {Wittmann},
  \citenamefont {Neumann}, \citenamefont {Jelezko}, \citenamefont {Rabeau},
  \citenamefont {Stavrias}, \citenamefont {Greentree}, \citenamefont {Prawer},
  \citenamefont {Meiler}, \citenamefont {Twamley}, \citenamefont {Hemmer},\
  and\ \citenamefont {Wrachtrup}}]{Gaebel2006}%
  \BibitemOpen
  \bibfield  {author} {\bibinfo {author} {\bibfnamefont {T.}~\bibnamefont
  {Gaebel}}, \bibinfo {author} {\bibfnamefont {M.}~\bibnamefont {Domhan}},
  \bibinfo {author} {\bibfnamefont {I.}~\bibnamefont {Popa}}, \bibinfo {author}
  {\bibfnamefont {C.}~\bibnamefont {Wittmann}}, \bibinfo {author}
  {\bibfnamefont {P.}~\bibnamefont {Neumann}}, \bibinfo {author} {\bibfnamefont
  {F.}~\bibnamefont {Jelezko}}, \bibinfo {author} {\bibfnamefont {J.~R.}\
  \bibnamefont {Rabeau}}, \bibinfo {author} {\bibfnamefont {N.}~\bibnamefont
  {Stavrias}}, \bibinfo {author} {\bibfnamefont {A.~D.}\ \bibnamefont
  {Greentree}}, \bibinfo {author} {\bibfnamefont {S.}~\bibnamefont {Prawer}},
  \bibinfo {author} {\bibfnamefont {J.}~\bibnamefont {Meiler}}, \bibinfo
  {author} {\bibfnamefont {J.}~\bibnamefont {Twamley}}, \bibinfo {author}
  {\bibfnamefont {P.~R.}\ \bibnamefont {Hemmer}}, \ and\ \bibinfo {author}
  {\bibfnamefont {J.}~\bibnamefont {Wrachtrup}},\ }\href {\doibase
  10.1038/nphys318} {\bibfield  {journal} {\bibinfo  {journal} {Nat. Phys.}\
  }\textbf {\bibinfo {volume} {2}},\ \bibinfo {pages} {408} (\bibinfo {year}
  {2006})}\BibitemShut {NoStop}%
\bibitem [{\citenamefont {Santori}\ \emph {et~al.}(2006)\citenamefont
  {Santori}, \citenamefont {Fattal}, \citenamefont {Spillane}, \citenamefont
  {Fiorentino}, \citenamefont {Beausoleil}, \citenamefont {Greentree},
  \citenamefont {Olivero}, \citenamefont {Draganski}, \citenamefont {Rabeau},
  \citenamefont {Reichart}, \citenamefont {Gibson}, \citenamefont {Rubanov},
  \citenamefont {Jamieson},\ and\ \citenamefont {Prawer}}]{Santori2006o}%
  \BibitemOpen
  \bibfield  {author} {\bibinfo {author} {\bibfnamefont {C.}~\bibnamefont
  {Santori}}, \bibinfo {author} {\bibfnamefont {D.}~\bibnamefont {Fattal}},
  \bibinfo {author} {\bibfnamefont {S.~M.}\ \bibnamefont {Spillane}}, \bibinfo
  {author} {\bibfnamefont {M.}~\bibnamefont {Fiorentino}}, \bibinfo {author}
  {\bibfnamefont {R.~G.}\ \bibnamefont {Beausoleil}}, \bibinfo {author}
  {\bibfnamefont {A.~D.}\ \bibnamefont {Greentree}}, \bibinfo {author}
  {\bibfnamefont {P.}~\bibnamefont {Olivero}}, \bibinfo {author} {\bibfnamefont
  {M.}~\bibnamefont {Draganski}}, \bibinfo {author} {\bibfnamefont {J.~R.}\
  \bibnamefont {Rabeau}}, \bibinfo {author} {\bibfnamefont {P.}~\bibnamefont
  {Reichart}}, \bibinfo {author} {\bibfnamefont {B.~C.}\ \bibnamefont
  {Gibson}}, \bibinfo {author} {\bibfnamefont {S.}~\bibnamefont {Rubanov}},
  \bibinfo {author} {\bibfnamefont {D.~N.}\ \bibnamefont {Jamieson}}, \ and\
  \bibinfo {author} {\bibfnamefont {S.}~\bibnamefont {Prawer}},\ }\href@noop {}
  {\bibfield  {journal} {\bibinfo  {journal} {Opt. Express}\ }\textbf {\bibinfo
  {volume} {14}},\ \bibinfo {pages} {7986} (\bibinfo {year}
  {2006})}\BibitemShut {NoStop}%
\bibitem [{\citenamefont {Waldermann}\ \emph {et~al.}(2007)\citenamefont
  {Waldermann}, \citenamefont {Olivero}, \citenamefont {Nunn}, \citenamefont
  {Surmacz}, \citenamefont {Wang}, \citenamefont {Jaksch}, \citenamefont
  {Taylor}, \citenamefont {Walmsley}, \citenamefont {Draganski}, \citenamefont
  {Reichart}, \citenamefont {Greentree}, \citenamefont {Jamieson},\ and\
  \citenamefont {Prawer}}]{Waldermann2007}%
  \BibitemOpen
  \bibfield  {author} {\bibinfo {author} {\bibfnamefont {F.~C.}\ \bibnamefont
  {Waldermann}}, \bibinfo {author} {\bibfnamefont {P.}~\bibnamefont {Olivero}},
  \bibinfo {author} {\bibfnamefont {J.}~\bibnamefont {Nunn}}, \bibinfo {author}
  {\bibfnamefont {K.}~\bibnamefont {Surmacz}}, \bibinfo {author} {\bibfnamefont
  {Z.~Y.}\ \bibnamefont {Wang}}, \bibinfo {author} {\bibfnamefont
  {D.}~\bibnamefont {Jaksch}}, \bibinfo {author} {\bibfnamefont {R.~A.}\
  \bibnamefont {Taylor}}, \bibinfo {author} {\bibfnamefont {I.~A.}\
  \bibnamefont {Walmsley}}, \bibinfo {author} {\bibfnamefont {M.}~\bibnamefont
  {Draganski}}, \bibinfo {author} {\bibfnamefont {P.}~\bibnamefont {Reichart}},
  \bibinfo {author} {\bibfnamefont {A.~D.}\ \bibnamefont {Greentree}}, \bibinfo
  {author} {\bibfnamefont {D.~N.}\ \bibnamefont {Jamieson}}, \ and\ \bibinfo
  {author} {\bibfnamefont {S.}~\bibnamefont {Prawer}},\ }\href {\doibase
  10.1016/j.diamond.2007.09.009} {\bibfield  {journal} {\bibinfo  {journal}
  {Diamond and Rel. Mat.}\ }\textbf {\bibinfo {volume} {16}},\ \bibinfo {pages}
  {1887} (\bibinfo {year} {2007})}\BibitemShut {NoStop}%
\bibitem [{\citenamefont {Maurer}\ \emph {et~al.}(2012)\citenamefont {Maurer},
  \citenamefont {Kucsko}, \citenamefont {Latta}, \citenamefont {Jiang},
  \citenamefont {Yao}, \citenamefont {Bennett}, \citenamefont {Pastawski},
  \citenamefont {Hunger}, \citenamefont {Chisholm}, \citenamefont {Markham},
  \citenamefont {Twitchen}, \citenamefont {Cirac},\ and\ \citenamefont
  {Lukin}}]{Maurer2012}%
  \BibitemOpen
  \bibfield  {author} {\bibinfo {author} {\bibfnamefont {P.~C.}\ \bibnamefont
  {Maurer}}, \bibinfo {author} {\bibfnamefont {G.}~\bibnamefont {Kucsko}},
  \bibinfo {author} {\bibfnamefont {C.}~\bibnamefont {Latta}}, \bibinfo
  {author} {\bibfnamefont {L.}~\bibnamefont {Jiang}}, \bibinfo {author}
  {\bibfnamefont {N.~Y.}\ \bibnamefont {Yao}}, \bibinfo {author} {\bibfnamefont
  {S.~D.}\ \bibnamefont {Bennett}}, \bibinfo {author} {\bibfnamefont
  {F.}~\bibnamefont {Pastawski}}, \bibinfo {author} {\bibfnamefont
  {D.}~\bibnamefont {Hunger}}, \bibinfo {author} {\bibfnamefont
  {N.}~\bibnamefont {Chisholm}}, \bibinfo {author} {\bibfnamefont
  {M.}~\bibnamefont {Markham}}, \bibinfo {author} {\bibfnamefont {D.~J.}\
  \bibnamefont {Twitchen}}, \bibinfo {author} {\bibfnamefont {J.~I.}\
  \bibnamefont {Cirac}}, \ and\ \bibinfo {author} {\bibfnamefont {M.~D.}\
  \bibnamefont {Lukin}},\ }\href {\doibase 10.1126/science.1220513} {\bibfield
  {journal} {\bibinfo  {journal} {Science}\ }\textbf {\bibinfo {volume}
  {336}},\ \bibinfo {pages} {1283} (\bibinfo {year} {2012})}\BibitemShut
  {NoStop}%
\bibitem [{\citenamefont {van~der Sar}\ \emph {et~al.}(2012)\citenamefont
  {van~der Sar}, \citenamefont {Wang}, \citenamefont {Blok}, \citenamefont
  {Bernien}, \citenamefont {Taminiau}, \citenamefont {Toyli}, \citenamefont
  {Lidar}, \citenamefont {Awschalom}, \citenamefont {Hanson},\ and\
  \citenamefont {Dobrovitski}}]{VanderSar2012}%
  \BibitemOpen
  \bibfield  {author} {\bibinfo {author} {\bibfnamefont {T.}~\bibnamefont
  {van~der Sar}}, \bibinfo {author} {\bibfnamefont {Z.~H.}\ \bibnamefont
  {Wang}}, \bibinfo {author} {\bibfnamefont {M.~S.}\ \bibnamefont {Blok}},
  \bibinfo {author} {\bibfnamefont {H.}~\bibnamefont {Bernien}}, \bibinfo
  {author} {\bibfnamefont {T.~H.}\ \bibnamefont {Taminiau}}, \bibinfo {author}
  {\bibfnamefont {D.~M.}\ \bibnamefont {Toyli}}, \bibinfo {author}
  {\bibfnamefont {D.~A.}\ \bibnamefont {Lidar}}, \bibinfo {author}
  {\bibfnamefont {D.~D.}\ \bibnamefont {Awschalom}}, \bibinfo {author}
  {\bibfnamefont {R.}~\bibnamefont {Hanson}}, \ and\ \bibinfo {author}
  {\bibfnamefont {V.~V.}\ \bibnamefont {Dobrovitski}},\ }\href {\doibase
  10.1038/nature10900} {\bibfield  {journal} {\bibinfo  {journal} {Nature
  (London)}\ }\textbf {\bibinfo {volume} {484}},\ \bibinfo {pages} {82}
  (\bibinfo {year} {2012})}\BibitemShut {NoStop}%
\bibitem [{\citenamefont {Neumann}\ and\ \citenamefont
  {Wrachtrup}(2013)}]{Neumann2013}%
  \BibitemOpen
  \bibfield  {author} {\bibinfo {author} {\bibfnamefont {P.}~\bibnamefont
  {Neumann}}\ and\ \bibinfo {author} {\bibfnamefont {J.}~\bibnamefont
  {Wrachtrup}},\ }\enquote {\bibinfo {title} {Optical engineering of
  diamond},}\ \ (\bibinfo  {publisher} {Wiley},\ \bibinfo {year} {2013})\
  Chap.~\bibinfo {chapter} {9}, p.\ \bibinfo {pages} {277}\BibitemShut
  {NoStop}%
\bibitem [{\citenamefont {Dolde}\ \emph {et~al.}(2013)\citenamefont {Dolde},
  \citenamefont {Jakobi}, \citenamefont {Naydenov}, \citenamefont {Zhao},
  \citenamefont {Pezzagna}, \citenamefont {Trautmann}, \citenamefont {Meijer},
  \citenamefont {Neumann}, \citenamefont {Jelezko},\ and\ \citenamefont
  {Wrachtrup}}]{Dolde2013}%
  \BibitemOpen
  \bibfield  {author} {\bibinfo {author} {\bibfnamefont {F.}~\bibnamefont
  {Dolde}}, \bibinfo {author} {\bibfnamefont {I.}~\bibnamefont {Jakobi}},
  \bibinfo {author} {\bibfnamefont {B.}~\bibnamefont {Naydenov}}, \bibinfo
  {author} {\bibfnamefont {N.}~\bibnamefont {Zhao}}, \bibinfo {author}
  {\bibfnamefont {S.}~\bibnamefont {Pezzagna}}, \bibinfo {author}
  {\bibfnamefont {C.}~\bibnamefont {Trautmann}}, \bibinfo {author}
  {\bibfnamefont {J.}~\bibnamefont {Meijer}}, \bibinfo {author} {\bibfnamefont
  {P.}~\bibnamefont {Neumann}}, \bibinfo {author} {\bibfnamefont
  {F.}~\bibnamefont {Jelezko}}, \ and\ \bibinfo {author} {\bibfnamefont
  {J.}~\bibnamefont {Wrachtrup}},\ }\href {\doibase 10.1038/nphys2545}
  {\bibfield  {journal} {\bibinfo  {journal} {Nature Physics}\ }\textbf
  {\bibinfo {volume} {9}},\ \bibinfo {pages} {139} (\bibinfo {year}
  {2013})}\BibitemShut {NoStop}%
\bibitem [{\citenamefont {Dolde}\ \emph {et~al.}(2014)\citenamefont {Dolde},
  \citenamefont {Bergholm}, \citenamefont {Wang}, \citenamefont {Jakobi},
  \citenamefont {Naydenov}, \citenamefont {Pezzagna}, \citenamefont {Meijer},
  \citenamefont {Jelezko}, \citenamefont {Neumann}, \citenamefont
  {Schulte-Herbr\"{u}ggen}, \citenamefont {Jacob},\ and\ \citenamefont
  {Wrachtrup}}]{Dolde2014}%
  \BibitemOpen
  \bibfield  {author} {\bibinfo {author} {\bibfnamefont {F.}~\bibnamefont
  {Dolde}}, \bibinfo {author} {\bibfnamefont {V.}~\bibnamefont {Bergholm}},
  \bibinfo {author} {\bibfnamefont {Y.}~\bibnamefont {Wang}}, \bibinfo {author}
  {\bibfnamefont {I.}~\bibnamefont {Jakobi}}, \bibinfo {author} {\bibfnamefont
  {B.}~\bibnamefont {Naydenov}}, \bibinfo {author} {\bibfnamefont
  {S.}~\bibnamefont {Pezzagna}}, \bibinfo {author} {\bibfnamefont
  {J.}~\bibnamefont {Meijer}}, \bibinfo {author} {\bibfnamefont
  {F.}~\bibnamefont {Jelezko}}, \bibinfo {author} {\bibfnamefont
  {P.}~\bibnamefont {Neumann}}, \bibinfo {author} {\bibfnamefont
  {T.}~\bibnamefont {Schulte-Herbr\"{u}ggen}}, \bibinfo {author} {\bibfnamefont
  {B.}~\bibnamefont {Jacob}}, \ and\ \bibinfo {author} {\bibfnamefont
  {J.}~\bibnamefont {Wrachtrup}},\ }\href {\doibase 10.1038/ncomms4371}
  {\bibfield  {journal} {\bibinfo  {journal} {Nature Communications}\ }\textbf
  {\bibinfo {volume} {5}},\ \bibinfo {pages} {3371} (\bibinfo {year}
  {2014})}\BibitemShut {NoStop}%
\bibitem [{\citenamefont {Pfaff}\ \emph {et~al.}()\citenamefont {Pfaff},
  \citenamefont {Hensen}, \citenamefont {Bernien}, \citenamefont {van Dam},
  \citenamefont {Blok}, \citenamefont {Taminiau}, \citenamefont {Tiggelman},
  \citenamefont {Schouten}, \citenamefont {Markham}, \citenamefont {Twitchen},\
  and\ \citenamefont {Hanson}}]{Pfaff2014}%
  \BibitemOpen
  \bibfield  {author} {\bibinfo {author} {\bibfnamefont {W.}~\bibnamefont
  {Pfaff}}, \bibinfo {author} {\bibfnamefont {B.}~\bibnamefont {Hensen}},
  \bibinfo {author} {\bibfnamefont {H.}~\bibnamefont {Bernien}}, \bibinfo
  {author} {\bibfnamefont {S.~B.}\ \bibnamefont {van Dam}}, \bibinfo {author}
  {\bibfnamefont {M.~S.}\ \bibnamefont {Blok}}, \bibinfo {author}
  {\bibfnamefont {T.~H.}\ \bibnamefont {Taminiau}}, \bibinfo {author}
  {\bibfnamefont {M.~J.}\ \bibnamefont {Tiggelman}}, \bibinfo {author}
  {\bibfnamefont {R.~N.}\ \bibnamefont {Schouten}}, \bibinfo {author}
  {\bibfnamefont {M.}~\bibnamefont {Markham}}, \bibinfo {author} {\bibfnamefont
  {D.~J.}\ \bibnamefont {Twitchen}}, \ and\ \bibinfo {author} {\bibfnamefont
  {R.}~\bibnamefont {Hanson}},\ }\href {\doibase 10.1126/science.1253512}
  {\bibinfo  {journal} {Science}\ ,\ \bibinfo {pages} {29 May 2014
  (10.1126/science.1253512)}}\BibitemShut {NoStop}%
\bibitem [{\citenamefont {Loubser}\ and\ \citenamefont
  {Van~Wyk}(1977)}]{Loubser1977}%
  \BibitemOpen
\bibfield  {journal} {  }\bibfield  {author} {\bibinfo {author} {\bibfnamefont
  {J.~H.~N.}\ \bibnamefont {Loubser}}\ and\ \bibinfo {author} {\bibfnamefont
  {J.~A.}\ \bibnamefont {Van~Wyk}},\ }\href@noop {} {\bibfield  {journal}
  {\bibinfo  {journal} {Diamond Research}\ }\textbf {\bibinfo {volume} {1}},\
  \bibinfo {pages} {11} (\bibinfo {year} {1977})}\BibitemShut {NoStop}%
\bibitem [{\citenamefont {Manson}\ \emph {et~al.}(2006)\citenamefont {Manson},
  \citenamefont {Harrison},\ and\ \citenamefont {Sellars}}]{Manson2006}%
  \BibitemOpen
  \bibfield  {author} {\bibinfo {author} {\bibfnamefont {N.~B.}\ \bibnamefont
  {Manson}}, \bibinfo {author} {\bibfnamefont {J.~P.}\ \bibnamefont
  {Harrison}}, \ and\ \bibinfo {author} {\bibfnamefont {M.~J.}\ \bibnamefont
  {Sellars}},\ }\href {\doibase 10.1103/PhysRevB.74.104303} {\bibfield
  {journal} {\bibinfo  {journal} {Phys. Rev. B}\ }\textbf {\bibinfo {volume}
  {74}},\ \bibinfo {pages} {104303} (\bibinfo {year} {2006})}\BibitemShut
  {NoStop}%
\bibitem [{\citenamefont {Delaney}\ \emph {et~al.}(2010)\citenamefont
  {Delaney}, \citenamefont {Greer},\ and\ \citenamefont
  {Larsson}}]{Delaney2010}%
  \BibitemOpen
  \bibfield  {author} {\bibinfo {author} {\bibfnamefont {P.}~\bibnamefont
  {Delaney}}, \bibinfo {author} {\bibfnamefont {J.~C.}\ \bibnamefont {Greer}},
  \ and\ \bibinfo {author} {\bibfnamefont {J.~A.}\ \bibnamefont {Larsson}},\
  }\href {\doibase 10.1021/nl903646p} {\bibfield  {journal} {\bibinfo
  {journal} {Nano Lett.}\ }\textbf {\bibinfo {volume} {10}},\ \bibinfo {pages}
  {610} (\bibinfo {year} {2010})}\BibitemShut {NoStop}%
\bibitem [{\citenamefont {van Oort}\ and\ \citenamefont
  {Glasbeek}(1989)}]{VanOort1989}%
  \BibitemOpen
  \bibfield  {author} {\bibinfo {author} {\bibfnamefont {E.}~\bibnamefont {van
  Oort}}\ and\ \bibinfo {author} {\bibfnamefont {M.}~\bibnamefont {Glasbeek}},\
  }\href@noop {} {\bibfield  {journal} {\bibinfo  {journal} {Phys. Rev. B}\
  }\textbf {\bibinfo {volume} {40}},\ \bibinfo {pages} {6509} (\bibinfo {year}
  {1989})}\BibitemShut {NoStop}%
\bibitem [{\citenamefont {Epstein}\ \emph {et~al.}(2005)\citenamefont
  {Epstein}, \citenamefont {Mendoza}, \citenamefont {Kato},\ and\ \citenamefont
  {Awschalom}}]{Epstein2005}%
  \BibitemOpen
  \bibfield  {author} {\bibinfo {author} {\bibfnamefont {R.~J.}\ \bibnamefont
  {Epstein}}, \bibinfo {author} {\bibfnamefont {F.~M.}\ \bibnamefont
  {Mendoza}}, \bibinfo {author} {\bibfnamefont {Y.~K.}\ \bibnamefont {Kato}}, \
  and\ \bibinfo {author} {\bibfnamefont {D.~D.}\ \bibnamefont {Awschalom}},\
  }\href {\doibase 10.1038/nphys141} {\bibfield  {journal} {\bibinfo  {journal}
  {Nat. Phys.}\ }\textbf {\bibinfo {volume} {1}},\ \bibinfo {pages} {94}
  (\bibinfo {year} {2005})}\BibitemShut {NoStop}%
\bibitem [{\citenamefont {Hanson}\ \emph
  {et~al.}(2006{\natexlab{b}})\citenamefont {Hanson}, \citenamefont {Mendoza},
  \citenamefont {Epstein},\ and\ \citenamefont {Awschalom}}]{Hanson2006}%
  \BibitemOpen
  \bibfield  {author} {\bibinfo {author} {\bibfnamefont {R.}~\bibnamefont
  {Hanson}}, \bibinfo {author} {\bibfnamefont {F.}~\bibnamefont {Mendoza}},
  \bibinfo {author} {\bibfnamefont {R.}~\bibnamefont {Epstein}}, \ and\
  \bibinfo {author} {\bibfnamefont {D.}~\bibnamefont {Awschalom}},\ }\href
  {\doibase 10.1103/PhysRevLett.97.087601} {\bibfield  {journal} {\bibinfo
  {journal} {Phys. Rev. Lett.}\ }\textbf {\bibinfo {volume} {97}},\ \bibinfo
  {pages} {08760} (\bibinfo {year} {2006}{\natexlab{b}})}\BibitemShut {NoStop}%
\bibitem [{\citenamefont {Rogers}\ \emph {et~al.}(2008)\citenamefont {Rogers},
  \citenamefont {Armstrong}, \citenamefont {Sellars},\ and\ \citenamefont
  {Manson}}]{Rogers2008}%
  \BibitemOpen
  \bibfield  {author} {\bibinfo {author} {\bibfnamefont {L.~J.}\ \bibnamefont
  {Rogers}}, \bibinfo {author} {\bibfnamefont {S.}~\bibnamefont {Armstrong}},
  \bibinfo {author} {\bibfnamefont {M.~J.}\ \bibnamefont {Sellars}}, \ and\
  \bibinfo {author} {\bibfnamefont {N.~B.}\ \bibnamefont {Manson}},\ }\href
  {\doibase 10.1088/1367-2630/10/10/103024} {\bibfield  {journal} {\bibinfo
  {journal} {New Journal of Physics}\ }\textbf {\bibinfo {volume} {10}},\
  \bibinfo {pages} {103024} (\bibinfo {year} {2008})}\BibitemShut {NoStop}%
\bibitem [{\citenamefont {Rogers}\ \emph {et~al.}(2009)\citenamefont {Rogers},
  \citenamefont {McMurtrie}, \citenamefont {Sellars},\ and\ \citenamefont
  {Manson}}]{Rogers2009}%
  \BibitemOpen
  \bibfield  {author} {\bibinfo {author} {\bibfnamefont {L.~J.}\ \bibnamefont
  {Rogers}}, \bibinfo {author} {\bibfnamefont {R.~L.}\ \bibnamefont
  {McMurtrie}}, \bibinfo {author} {\bibfnamefont {M.~J.}\ \bibnamefont
  {Sellars}}, \ and\ \bibinfo {author} {\bibfnamefont {N.~B.}\ \bibnamefont
  {Manson}},\ }\href {\doibase 10.1088/1367-2630/11/6/063007} {\bibfield
  {journal} {\bibinfo  {journal} {New Journal of Physics}\ }\textbf {\bibinfo
  {volume} {11}},\ \bibinfo {pages} {063007} (\bibinfo {year}
  {2009})}\BibitemShut {NoStop}%
\bibitem [{\citenamefont {Lai}\ \emph {et~al.}(2009)\citenamefont {Lai},
  \citenamefont {Zheng}, \citenamefont {Jelezko}, \citenamefont {Treussart},\
  and\ \citenamefont {Roch}}]{Lai2009}%
  \BibitemOpen
  \bibfield  {author} {\bibinfo {author} {\bibfnamefont {N.}~\bibnamefont
  {Lai}}, \bibinfo {author} {\bibfnamefont {D.}~\bibnamefont {Zheng}}, \bibinfo
  {author} {\bibfnamefont {F.}~\bibnamefont {Jelezko}}, \bibinfo {author}
  {\bibfnamefont {F.}~\bibnamefont {Treussart}}, \ and\ \bibinfo {author}
  {\bibfnamefont {J.-F.}\ \bibnamefont {Roch}},\ }\href {\doibase
  10.1063/1.3238467} {\bibfield  {journal} {\bibinfo  {journal} {Appl. Phys.
  Lett.}\ }\textbf {\bibinfo {volume} {95}},\ \bibinfo {pages} {133101}
  (\bibinfo {year} {2009})}\BibitemShut {NoStop}%
\bibitem [{\citenamefont {Armstrong}\ \emph {et~al.}(2010)\citenamefont
  {Armstrong}, \citenamefont {Rogers}, \citenamefont {Mcmurtrie},\ and\
  \citenamefont {Manson}}]{Armstrong2010}%
  \BibitemOpen
  \bibfield  {author} {\bibinfo {author} {\bibfnamefont {S.}~\bibnamefont
  {Armstrong}}, \bibinfo {author} {\bibfnamefont {L.~J.}\ \bibnamefont
  {Rogers}}, \bibinfo {author} {\bibfnamefont {R.~L.}\ \bibnamefont
  {Mcmurtrie}}, \ and\ \bibinfo {author} {\bibfnamefont {N.~B.}\ \bibnamefont
  {Manson}},\ }\href {\doibase 10.1016/j.phpro.2010.01.223} {\bibfield
  {journal} {\bibinfo  {journal} {Physics Procedia}\ }\textbf {\bibinfo
  {volume} {3}},\ \bibinfo {pages} {1569} (\bibinfo {year} {2010})}\BibitemShut
  {NoStop}%
\bibitem [{\citenamefont {Hanle}(1924)}]{Hanle1924}%
  \BibitemOpen
  \bibfield  {author} {\bibinfo {author} {\bibfnamefont {W.}~\bibnamefont
  {Hanle}},\ }\href {\doibase 10.1007/BF01331827} {\bibfield  {journal}
  {\bibinfo  {journal} {Z. Phys.}\ }\textbf {\bibinfo {volume} {30}},\ \bibinfo
  {pages} {93} (\bibinfo {year} {1924})}\BibitemShut {NoStop}%
\bibitem [{\citenamefont {Woodward}(2010)}]{Woodward2010}%
  \BibitemOpen
  \bibfield  {author} {\bibinfo {author} {\bibfnamefont {J.~R.}\ \bibnamefont
  {Woodward}},\ }\enquote {\bibinfo {title} {Carbon-centered free radicals and
  radical cations: Structure, reactivity, and dynamics},}\ \ (\bibinfo
  {publisher} {Wiley},\ \bibinfo {year} {2010})\ Chap.~\bibinfo {chapter} {8},
  p.\ \bibinfo {pages} {157}\BibitemShut {NoStop}%
\bibitem [{\citenamefont {Anisimov}\ \emph {et~al.}(1983)\citenamefont
  {Anisimov}, \citenamefont {Grigoryants}, \citenamefont {Kiyanov},
  \citenamefont {Salikhov}, \citenamefont {Sukhenko},\ and\ \citenamefont
  {Molin}}]{Anisimov1983}%
  \BibitemOpen
  \bibfield  {author} {\bibinfo {author} {\bibfnamefont {O.~A.}\ \bibnamefont
  {Anisimov}}, \bibinfo {author} {\bibfnamefont {V.~M.}\ \bibnamefont
  {Grigoryants}}, \bibinfo {author} {\bibfnamefont {S.~V.}\ \bibnamefont
  {Kiyanov}}, \bibinfo {author} {\bibfnamefont {K.~M.}\ \bibnamefont
  {Salikhov}}, \bibinfo {author} {\bibfnamefont {S.~A.}\ \bibnamefont
  {Sukhenko}}, \ and\ \bibinfo {author} {\bibfnamefont {Y.~N.}\ \bibnamefont
  {Molin}},\ }\href {\doibase 10.1007/BF00519845} {\bibfield  {journal}
  {\bibinfo  {journal} {Theor. Exp. Chem.}\ }\textbf {\bibinfo {volume} {18}},\
  \bibinfo {pages} {256} (\bibinfo {year} {1983})}\BibitemShut {NoStop}%
\bibitem [{\citenamefont {Fischer}(1983)}]{Fischer1983}%
  \BibitemOpen
  \bibfield  {author} {\bibinfo {author} {\bibfnamefont {H.}~\bibnamefont
  {Fischer}},\ }\href {\doibase 10.1016/0009-2614(83)87287-X} {\bibfield
  {journal} {\bibinfo  {journal} {Chem. Phys. Lett.}\ }\textbf {\bibinfo
  {volume} {100}},\ \bibinfo {pages} {255} (\bibinfo {year}
  {1983})}\BibitemShut {NoStop}%
\bibitem [{\citenamefont {Zaitsev}(2001)}]{Zaitsev2001}%
  \BibitemOpen
  \bibfield  {author} {\bibinfo {author} {\bibfnamefont {A.~M.}\ \bibnamefont
  {Zaitsev}},\ }\href@noop {} {\emph {\bibinfo {title} {Optical propaties of
  diamond: a data of handbook}}}\ (\bibinfo  {publisher} {Springer},\ \bibinfo
  {year} {2001})\BibitemShut {NoStop}%
\bibitem [{\citenamefont {Lawson}\ \emph {et~al.}(1998)\citenamefont {Lawson},
  \citenamefont {Fisher}, \citenamefont {Hunt},\ and\ \citenamefont
  {Newton}}]{Lawson1998}%
  \BibitemOpen
  \bibfield  {author} {\bibinfo {author} {\bibfnamefont {S.~C.}\ \bibnamefont
  {Lawson}}, \bibinfo {author} {\bibfnamefont {D.}~\bibnamefont {Fisher}},
  \bibinfo {author} {\bibfnamefont {D.~C.}\ \bibnamefont {Hunt}}, \ and\
  \bibinfo {author} {\bibfnamefont {M.~E.}\ \bibnamefont {Newton}},\
  }\href@noop {} {\bibfield  {journal} {\bibinfo  {journal} {J. Phys.: Condens.
  Matter}\ }\textbf {\bibinfo {volume} {10}},\ \bibinfo {pages} {6171}
  (\bibinfo {year} {1998})}\BibitemShut {NoStop}%
\bibitem [{\citenamefont {Nadolinny}\ and\ \citenamefont
  {Yelisseyev}(1994)}]{Nadolinny1994}%
  \BibitemOpen
  \bibfield  {author} {\bibinfo {author} {\bibfnamefont {V.~A.}\ \bibnamefont
  {Nadolinny}}\ and\ \bibinfo {author} {\bibfnamefont {A.~P.}\ \bibnamefont
  {Yelisseyev}},\ }\href {\doibase 10.1016/0925-9635(94)90024-8} {\bibfield
  {journal} {\bibinfo  {journal} {Diamond and Rel. Mat.}\ }\textbf {\bibinfo
  {volume} {3}},\ \bibinfo {pages} {17} (\bibinfo {year} {1994})}\BibitemShut
  {NoStop}%
\bibitem [{\citenamefont {Yelisseyev}\ \emph {et~al.}(1996)\citenamefont
  {Yelisseyev}, \citenamefont {Nadolinny}, \citenamefont {Feigelson},
  \citenamefont {Terentyev},\ and\ \citenamefont {S}}]{Yelisseyev1996}%
  \BibitemOpen
  \bibfield  {author} {\bibinfo {author} {\bibfnamefont {A.}~\bibnamefont
  {Yelisseyev}}, \bibinfo {author} {\bibfnamefont {V.}~\bibnamefont
  {Nadolinny}}, \bibinfo {author} {\bibfnamefont {B.}~\bibnamefont
  {Feigelson}}, \bibinfo {author} {\bibfnamefont {S.}~\bibnamefont
  {Terentyev}}, \ and\ \bibinfo {author} {\bibfnamefont {N.}~\bibnamefont
  {S}},\ }\href {\doibase 10.1016/0925-9635(96)00511-0} {\bibfield  {journal}
  {\bibinfo  {journal} {Diamond and Rel. Mat.}\ }\textbf {\bibinfo {volume}
  {5}},\ \bibinfo {pages} {1113} (\bibinfo {year} {1996})}\BibitemShut
  {NoStop}%
\bibitem [{\citenamefont {Siyushev}\ \emph {et~al.}(2013)\citenamefont
  {Siyushev}, \citenamefont {Pinto}, \citenamefont {V\"or\"os}, \citenamefont
  {Gali}, \citenamefont {Jelezko},\ and\ \citenamefont
  {Wrachtrup}}]{Siyushev2013}%
  \BibitemOpen
  \bibfield  {author} {\bibinfo {author} {\bibfnamefont {P.}~\bibnamefont
  {Siyushev}}, \bibinfo {author} {\bibfnamefont {H.}~\bibnamefont {Pinto}},
  \bibinfo {author} {\bibfnamefont {M.}~\bibnamefont {V\"or\"os}}, \bibinfo
  {author} {\bibfnamefont {A.}~\bibnamefont {Gali}}, \bibinfo {author}
  {\bibfnamefont {F.}~\bibnamefont {Jelezko}}, \ and\ \bibinfo {author}
  {\bibfnamefont {J.}~\bibnamefont {Wrachtrup}},\ }\href {\doibase
  10.1103/PhysRevLett.110.167402} {\bibfield  {journal} {\bibinfo  {journal}
  {Phys. Rev. Lett.}\ }\textbf {\bibinfo {volume} {110}},\ \bibinfo {pages}
  {167402} (\bibinfo {year} {2013})}\BibitemShut {NoStop}%
\bibitem [{\citenamefont {Kennedy}\ \emph {et~al.}(2003)\citenamefont
  {Kennedy}, \citenamefont {Colton}, \citenamefont {Butler}, \citenamefont
  {Linares},\ and\ \citenamefont {Doering}}]{Kennedy2003}%
  \BibitemOpen
  \bibfield  {author} {\bibinfo {author} {\bibfnamefont {T.~A.}\ \bibnamefont
  {Kennedy}}, \bibinfo {author} {\bibfnamefont {J.~S.}\ \bibnamefont {Colton}},
  \bibinfo {author} {\bibfnamefont {J.~E.}\ \bibnamefont {Butler}}, \bibinfo
  {author} {\bibfnamefont {R.~C.}\ \bibnamefont {Linares}}, \ and\ \bibinfo
  {author} {\bibfnamefont {P.~J.}\ \bibnamefont {Doering}},\ }\href {\doibase
  10.1063/1.1626791} {\bibfield  {journal} {\bibinfo  {journal} {Appl. Phys.
  Lett.}\ }\textbf {\bibinfo {volume} {83}},\ \bibinfo {pages} {4190} (\bibinfo
  {year} {2003})}\BibitemShut {NoStop}%
\end{thebibliography}%

\end{document}